\documentstyle[aps,multicol,epsf]{revtex}
\draft
\def\beginwide{
        \end{multicols} \vspace*{-0.5cm} \noindent
        \rule{3.5in}{.1mm}\rule{.1mm}{5mm} \widetext \medskip }
\def\beginwidetop{
        \end{multicols} \vspace*{-0.5cm} \noindent
        \widetext \medskip }
\def\endwide{
        \hspace*{3.35in}~\rule[-5mm]{.1mm}{5mm}\rule{3.5in}{.1mm}
        \begin{multicols}{2} \vspace*{-1.0cm} \noindent }
\def\endwidebottom{
        \begin{multicols}{2} \vspace*{-1.0cm} \noindent }
 
\newcommand{\beq}{\begin{equation}}
\newcommand{\eeq}{\end{equation}}
\newcommand{\bdis}{\begin{displaymath}}
\newcommand{\edis}{\end{displaymath}}
\newcommand{\bea}{\begin{eqnarray}}
\newcommand{\eea}{\end{eqnarray}}
\newcommand{\barr}{\begin{array}}
\newcommand{\earr}{\end{array}}
\begin{document}
 
\title{Mean-field behavior of the sandpile model below the upper 
critical dimension}

\author{Alessandro Chessa$^{(1,2)}$, Enzo Marinari$^{(1,3)}$, 
Alessandro Vespignani$^{(4)}$ and Stefano Zapperi$^{(5)}$}
 
\address{
1)Dipartimento di Fisica, Universit\'a di Cagliari, 
Via Ospedale 72, 09124 Cagliari, Italy\\
2)Istituto Nazionale di Fisica della Materia (INFM), Sezione di 
Cagliari, Italy\\
3)Istituto Nazionale di Fisica Nucleare (INFN), Sezione di Cagliari, 
Italy\\
4)International Centre for Theoretical Physics (ICTP), 
P.O. Box 586, 34100 Trieste, Italy\\
5)Center for Polymer Studies and Department of Physics
Boston University, Boston, MA 02215}
 
\date{\today}
 
\maketitle
\begin{abstract}
We present results of large scale numerical simulations of the Bak, Tang
and 
Wiesenfeld sandpile model. We analyze the critical behavior of the model 
in Euclidean dimensions $2\leq d\leq 6$. We consider  
a dissipative generalization of the model and study the 
avalanche size and duration distributions for different values of
the lattice size and dissipation.
We find that the scaling exponents 
in $d=4$ significantly differ from mean-field predictions,
thus suggesting  an upper critical dimension $d_c\geq 5$. 
Using the relations among the dissipation rate $\epsilon$ and the 
finite lattice size $L$,
we find that a subset of the exponents displays mean-field 
values below the upper critical dimensions. This behavior 
is explained in terms of conservation laws.
\end{abstract}
 
 
%
%

\begin{multicols}{2}
Since the introduction of the concept of self-organized criticality
(SOC) 
ten years ago \cite{btw,grin}, an enormous effort has been devoted to
the 
understanding of this irreversible dynamical phenomenon. 
SOC models  oppose the standard picture
of critical phenomena, since their dynamics should
generate a self-organization of the system into a critical state,
without 
need for the fine tuning of external parameters. 
The paradigmatic SOC model 
is the sandpile automaton, in which a slow external driving 
of sand particles leads to a stationary state with avalanches
distributed on 
all length scales \cite{btw}. Despite the apparently simple rules, the
model 
shows a complicated behavior which is not amenable to a complete
solution.

In SOC models, the concept of ``spontaneous'' criticality is 
quite ambiguous because  
it has been recognized that criticality appears only if the 
driving rate is fine tuned to zero \cite{grin,sorn,vz}. 
The slow driving assumption implies 
nonlocality in the dynamical rules of the model \cite{nonloc}, which
makes a general theory of SOC problematic \cite{theo}. 
Several important theoretical questions are 
still not resolved, such as 
the precise definition of universality classes, 
the value of the upper critical dimension,
and the validity of fluctuation-dissipation theorems. 
These problems are also reflected in the relatively few exact results 
available in the literature \cite{dhar,exact}.
Furthermore, these issues are also unclear from the 
numerical point of view, and only in the last years, earlier 
computational efforts\cite{grasma,manna1} have been followed by 
more accurate numerical studies \cite{ben,lubeck,lubeck2}. 

Recently, a general dynamical mean-field (MF) analysis\cite{vz} of 
sandpile models pointed out the similarities between SOC models 
and phase transitions in systems with absorbing states\cite{dck}.
Criticality 
is analyzed in terms of the response function singularities and the MF 
critical exponents are calculated.
This  method relates bulk and boundary dissipation and introduces a 
scaling 
relation relating dissipation and finite-size effects. Moreover, due to
the 
conservative nature of sandpiles at the critical point, 
a subset of critical exponent was predicted to 
display MF values also in low dimensions \cite{vz}.
This result provides an important test 
to verify the validity of the MF theory, 
and can be used as a consistency check in the numerical analysis 
of several exponents characterizing sandpile models.

Here, we study the critical behavior of the avalanche 
size and duration distribution in order to provide numerical 
evidences for the MF behavior of low dimensional sandpiles.
We perform an  accurate study of critical exponents   for 
conservative \cite{btw}
and dissipative \cite{manna2,chris} 
sandpiles in dimensionality ranging from $d=2$ to $d=6$. 
This allows us to estimate 
the upper critical dimension $d_c$. In contrast with recent 
numerical simulations \cite{lubeck2},
MF behavior is observed only in $d=6$ and we therefore 
exclude that $d_c=4$.
In addition we found that  some  critical exponents assume
constantly their MF values in all dimensions $d$, 
as predicted in Ref.~\cite{vz}.

We consider the $d$-dimensional Bak, Tang and Wiesenfeld (BTW) 
sandpile model\cite{btw} on  a hypercubic lattice of size $L$.  
On each site $i$ of the lattice we define an integer variable 
$z_i$ which is identified with the sand or energy stored in the site. 
At each time step an energy grain is added on a randomly chosen site
($z_i\to z_i+1$). When one of the sites reaches or exceed the threshold
$z_c=2d$ a dynamical process occurs: $z_i=z_i-2d$ and $z_j=z_j+1$, where 
$j$ represents the nearest neighbor sites. Such a ``toppling'' event 
can induce nearest neighbor sites to topple on 
their turn and so on, until all sites are below the critical threshold.
This process is called an avalanche. The slow driving condition is 
implemented by stopping the random energy addition during the avalanche 
spreading. This means that the driving time scale is infinitely slow
with 
respect to the avalanche characteristic time. 

The model is locally conservative; no energy grains are lost during the 
toppling event. The only dissipation occurs at the boundary, from 
which energy can leave the system. We also use a nonconservative
definition 
of the model. With probability $p$ the toppling site loose its energy
without transferring it to its nearest neighbors. This means that on
average
a quantity $\epsilon=2d p$ of energy is dissipated in each toppling. In
this 
case periodic boundary conditions can be considered. With both these 
definitions, 
the model reaches a stationary state in which the energy introduced 
by the external random drive is balanced on average by the energy
dissipated 
in the dynamical evolution.
In the stationary state, we can define the probability that 
the addition of a single grain is followed by an avalanche of 
$s$ relaxation events. In the limit 
$\epsilon\to 0$, it is possible to show that the system response
function is diverging, revealing 
the presence a critical point \cite{vz}.  Close to 
criticality, the avalanche size 
distribution assumes the scaling form
\beq 
P(s)=s^{-\tau}{\cal G}(s/s_c)\ ,
\eeq
where $s_c$ is the cutoff in the avalanche size. 

In the infinite time scale separation, the cut-off size is a function 
$s_c\sim\epsilon^{-1/\sigma}$ of the bulk or 
border dissipation. The boundary dissipation follows the scaling 
form $\epsilon\sim L^{-\mu}$, where $\mu$ is the  
exponent relating the dissipation rate with the system size. 
Thus we obtain that in the case of a fully conservative system 
$s_c\sim L^{\mu/\sigma}$. It is useful to introduce also the avalanche 
characteristic length $\xi$ and the scaling relations $s_c\sim\xi^D$ and 
$\xi\sim\epsilon^{-\nu}$, which define the the fractal dimension and the 
characteristic length divergence exponents, respectively.  
By noting that $\xi$ and $L$ must rescale in the same way,  we
immediately 
obtain the scaling relations:
\beq
D\sigma=\nu^{-1},~~~\nu=\mu^{-1}.
\label{scal1}
\eeq
The MF theory gives $\tau_{MF}=3/2$, $\sigma_{MF}=1/2$ and 
$D_{MF}=4$\cite{vz}. In addition, the theory of Ref.~\cite{vz} 
predicts that $\mu=2$ and $\nu=1/2$ in all dimensions because of 
the inherent conservation law of these models. 
The values of these two exponents also imply that $<s>\sim L^2$ and 
$<s>\sim\epsilon^{-\gamma}$ with $\gamma=1$ for any $d$ \cite{nota1}.  
{}From these results, we obtain the scaling relation $D\sigma=2$, that
also holds for all $d$. These   
results provide a powerful consistency check 
in the numerical analysis of several 
exponents characterizing sandpile models.
The value of the exponents $\tau,\sigma$ and $D$ depend on $d$ 
and will only agree with 
MF theory values when $d>d_c$. 

In order to test the above picture we have studied the 
avalanche size distribution 
in systems with dimension ranging from $d=2$ to $d=6$ and varying sizes
$L$ 
and dissipation $\epsilon$. In the first simulation set ($\epsilon=0$), 
system sizes $L\leq 1024$ for $d=2$, $L\leq 762$ for $d=3$, 
$L\leq 144$ for $d=4$, $L\leq 53$ for $d=5$ and $L\leq 27$ for $d=6$
have been investigated. 
In the second set the dissipation rates change with the dimension: 
$\epsilon \geq 10^{-5}$ for $d=2$, $\epsilon \geq 10^{-4}$ for $d=3$ and
$\epsilon \geq 10^{-1}$ for $d=4,5$ and $6$
with lattice of the maximum size available.
In each case, statistical distributions are obtained averaging over 
a number ranging from  $10^6$ to $10^7$ nonzero avalanches. 
For $d\geq 3$, the sizes reached in our simulations 
are, to our knowledge, the largest which have ever been used. 
In $d=2$ we did not push the computational effort too far, since 
this case is studied in the literature also for very large 
lattice sizes \cite{lubeck}. 
Particular attention must be paid in performing simulations with 
dissipation, because if the dissipation is too small,  
$\xi$ can become larger than $L$ leading to spurious results for the
cutoff. It is easy to recognize that diminishing the 
dissipation rates is similar to increasing the system sizes; in both
cases the  average avalanche size is increasing.

Our simulations provide two independent estimates of the 
exponent $\tau$ by extrapolating the power law behavior 
for different sizes $L$ and finite dissipation rates.  
The numerical determination of 
an overall power law behavior determined with a ten percent 
accuracy is an easy task.
On the contrary, to increase the accuracy of an order of magnitude
requires 
a very careful data treatment. 
We noticed that the individuation  of 
the straight portion of the probability distribution  is a very 
delicate point in the accurate 
evaluation of the exponent $\tau$. In particular, even innocuous
smoothing
procedures give rise to impressive systematic bias. 
In fact, the fit of the exponent $\tau$ suffers from strong systematic
errors 
due to the lower and upper cutoff of the distribution. 
For this reason, we perform
a local slope analysis of the raw data 
by studying the behavior of the logarithmic derivative 
of each avalanche distribution. In this way, it is possible to identify 
a plateau in which the local slope is almost constant. This plateau
defines 
the range of $s$  we can use for a meaningful determination of the
exponent 
$\tau$. Naturally, this range is increasing for larger sizes $L$ and 
smaller dissipation rates $\epsilon$. 
Nevertheless, the  measurements of $\tau$ presents
strong finite size effects especially in  
$d=2$. In this case the exponent $\tau$ seems to suffer from 
logarithmic corrections with the size $L$; i.e. $\tau(L)=\tau
-const/\log L$.
In $d>2$, the numerical evidences show a much faster convergence
estimated
as $\tau(L)=\tau -const L^{-2}$. In the literature, the asymptotic
estimates
of $\tau$ are obtained through extrapolation from the previous
functional 
behavior\cite{grasma,manna1,lubeck,lubeck2}.
For greater accuracy we used also a new 
extrapolation procedure  devised in Ref.\cite{lubeck}.
This procedure improves the determination of the exponent by using the 
functional form of the corrections for the direct determination of
$\tau$
by comparing different size samples. 
In table I we report the asymptotic values of the exponent $\tau$ for 
$2\leq d\leq 5$. The values are in good agreement with previous estimate
from
ref.s\cite{grasma,manna1,lubeck}.
In addition, it appears from the results of table I that also in $d\geq4$
the measured value is not definitely converged on the MF result.
The values extrapolated in presence of finite dissipation rates
$\epsilon$
have a small systematic discrepancy with respect to the values obtained 
in the usual extrapolation procedure. However this is to be ascribed to 
the different boundary conditions used in the simulations. 
It is worth to remark that, as already pointed out by other 
authors \cite{ben}, the sole analysis of 
$\tau$ can be misleading, 
since this exponent is not very sensible to the variations of the
dimension $d$, as well as variations of universality class \cite{ben}.
The exponent, in fact, suffers a maximum variation of around 10 per cent 
with respect to its MF value.
The simple analysis of this exponent is therefore not always determinant
in 
the discrimination of many of the crucial properties of sandpile models.

In order to provide another independent estimate of the exponents
$\tau$,
$D$ and $\sigma$, we perform a data collapse analysis, which
turns out to be very powerful in this case.
Under the finite size scaling assumptions, 
the distributions $P(s,L)$
and $P(s,\epsilon)$ collapse onto a single curve 
if we rescale properly the variables.
Thus, by defining $P_{q_x}=P(s,x)/s^{-\tau}$ and $q_L=sL^{-D}$ 
($q_\epsilon=s\epsilon^{1/\sigma}$) we obtain that all data must
collapse
onto the universal function:
\beq 
P_{q_x}={\cal G}(q_x).
\eeq
The exponent $\tau$ controls  the rescaling of the vertical axis,
while the exponents $D$ and $\sigma$ define the rescaling of the
horizontal 
axis. Similar universal function can be obtained by using as rescaling
variables
$L$ or $\epsilon$, thus obtaining $P(s,x)L^{D\tau}={\cal F}(sL^{-D})$
and 
$P(s,x)\epsilon^{-\tau/\sigma}={\cal H}(s\epsilon^{1/\sigma})$
The same analysis can be also performed on the integrated distribution 
$P(s*>s)$, that is usually less noisy. In this case the power law
behavior 
is governed by the exponent $\tau -1$. In order to test carefully the 
numerical data  we repeated the data collapse analysis by using all 
the previous data collapse forms as well as a direct fitting procedure. 
We show in figure 1 and 2 the data collapse for the conservative and 
dissipative BTW model in $d=4$. We obtain 
very precise collapses, which are very sensible to the tuning of the 
the various exponents. The evaluation of  exponents by a direct fit 
obtains results which are in perfect agreement with the data collapse 
analysis. In table II we report the values of the various exponents in 
$2\leq d\leq 5$. 
{}From the present analysis, we verify that
$\nu^{-1}=D\sigma\simeq 2.0$
independently on the dimension. 
As expected also the exponent governing the divergence of 
the average size assumes constantly the  value $\gamma\simeq 1$.
It is a striking evidence that while 
the exponents $D$ and $\sigma$ vary from $d=2$ to $d=5$ more than 
$30$ percent with a clear trend to the MF values, 
their product fluctuates of just a few percent. This definitely shows
that 
the dynamics of sandpile maintains MF features also in low dimensions as
shown in ref.\cite{vz}. Furthermore, the constant value of 
$\nu^{-1}=D\sigma$ provides an additional consistency check for
reliability of our results.
 
Looking at table II, we see a strong indication that MF behavior has
not yet set in $d=4$. In fact, contrary to some recent numerical 
results \cite{lubeck2}, we obtain that $D\simeq3.5$ and that 
$\sigma^{-1}\simeq 1.7$. These values, obtained by data collapse,  
are undoubtedly far from the MF ones. They are also fully
compatible with the exponent $\tau$ as measured independently from the
extrapolation procedure. 
In fact, $\gamma,\sigma$ and $\tau$ have to satisfy 
the scaling relations $\sigma\gamma=2-\tau$ \cite{vz}, which is fully 
consistent with the measured values. For these reasons, we are confident
in excluding that $d=4$ is the upper critical dimension of the sandpile
model. 

In order to provide a further check to the previous results we also
analyzed the avalanche duration distributions.  
The results that will appear in a forthcoming paper\cite{cmvz}, 
confirms the scenario presented in this letter.
Here, we only report the results 
concerning $d=4$, which are important being discriminating for the upper 
critical dimension. By using the data collapse 
described previously we measured 
the dynamical critical exponents $t_c\sim L^z$ and  
$t_c\sim\epsilon^{-\Delta}$ 
defining the divergence of the 
characteristic time $t_c$ with respect to the system size and 
dissipation rate, respectively. In $d=4$ we obtain 
$z=1.8\pm 0.1$ and $\Delta=0.90\pm 0.05$, that also in this case are
different 
from the MF values  $z_{MF}=2$ and $\Delta_{MF}=1$. This again supports
the claim that $d_c>4$. 

The value of the upper critical dimension is a long standing 
theoretical question 
in the study of sandpile models. 
Several theoretical estimates (none of them rigorous) give
$d_c=4$\cite{theo},
that has been obtained also from recent numerical simulations
\cite{lubeck2}. 
In contrast, other numerical studies \cite{chris} and the analogies with 
dynamical percolation led several authors to conjecture $d_c=6$. 
>From the analysis of our data, that have been obtained using
the largest lattice sizes ever used, we can definitely say that $d_c>4$.
In $d=5$ we note discrepancies between the values
we measure and MF predictions. However, because of 
the relatively small sizes reached in this case, we can not exclude that 
deviations from the MF behavior are due to finite size effects. In $d=6$
we 
obtain the MF values, but the error bars do not permit a reliable 
discussion of the results. 

The main part of the numerical simulations have been run on the {\em
Kalix} parallel computer \cite{kalix} (a Beowulf project at Cagliari
Physics Department). We thank G. Mula for leading the effort
toward organizing this computer facility. The Center for Polymer Studies
is supported by NSF.

\begin{narrowtext}
\begin{table}[t]
\begin{tabular}{|c|c|c|c|c|}
\hline
d & $2$ & $3$ & $4$ & $5$\\
\hline \hline
$\tau_\infty^{L}$ & $1.30\pm0.01$ & $1.33\pm0.01$ 
& $1.45\pm0.01$ & $1.51\pm0.01$ \\
$\tau_\infty^{\epsilon}$ & $1.25\pm0.03$ & $1.31\pm0.01$ & 
$1.43\pm0.01$ & $1.49\pm0.01$ \\
\hline
\end{tabular} 
\caption{Exponent $\tau^{L}$ and $\tau^{\epsilon}$
obtained from different sizes and dissipation rates extrapolation, 
respectively. The systematic difference is given by the different 
boundary conditions used in the numerical simulations.}
\label{tab1}
\end{table}
\end{narrowtext}

\begin{narrowtext}
\begin{table}[t]
\begin{tabular}{|c|c|c|c|c|}
\hline
d & $\gamma$ & $D$ & $1/\sigma$ & $\nu=(D\sigma)^{-1}$\\
\hline \hline
2 & $0.998\pm0.001$ & $2.7\pm0.1$ & $1.30\pm0.05$ & $0.48 \pm0.03 $ \\
3 & $0.998\pm0.001$ & $3.0\pm0.1$ & $1.50\pm0.05$ & $0.50 \pm0.02 $ \\
4 & $0.978\pm0.003$ & $3.5\pm0.1$ & $1.72\pm0.05$ & $0.49 \pm0.02 $ \\
5 & $0.990\pm0.003$ & $3.8\pm0.1$ & $1.88\pm0.05$ & $0.50 \pm0.02 $ \\
\hline
\end{tabular} 
\caption{Values of the critical exponents in different dimensions. 
The results obtained with data collapse analysis and direct 
fitting procedure are the same and the errors we report is the one 
estimated by the fitting procedure. }
\label{tab2}
\end{table}
\end{narrowtext}
\begin{figure}[htb]
\narrowtext
\centerline{
        \epsfxsize=7.0cm
        \epsfbox{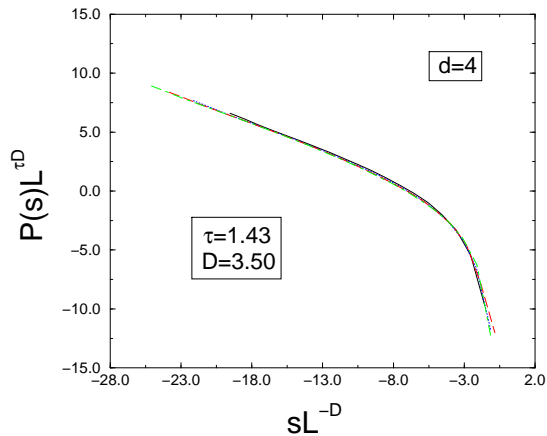}
        }
\caption{Data collapse analysis of the avalanche size distribution 
of the BTW model in $d=4$ with $\epsilon\equiv 0$ and for lattice sizes
$L=48,80,112$ and $144$. The plot is on a double logarithmic scale.}
\label{fig:2}
\end{figure}
\begin{figure}[htb]
\narrowtext
\centerline{
        \epsfxsize=7.0cm
        \epsfbox{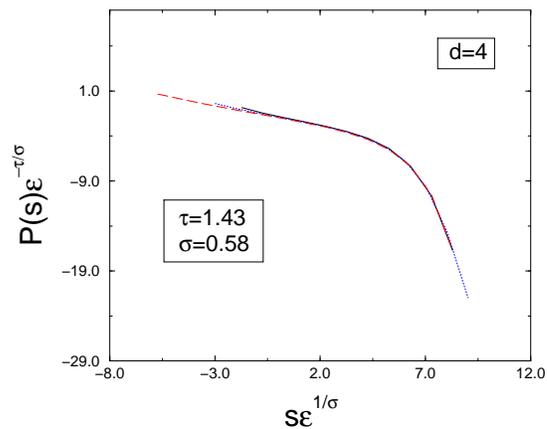}
        }
\caption{Data collapse analysis of the avalanche size distribution 
of the BTW model in $d=4$ with $L=144$ and for dissipation rates 
$\epsilon>10^{-1}$.The plot is on a double logarithmic scale. }
\label{fig:3}
\end{figure}

\end{multicols}

\begin{thebibliography}{99}
\bibitem{nonloc}
        Sandpile models are driven by 
        adding a single energy grain on a 
        randomly chosen site, when no active sites are present. 
        In this way, avalanches are instantaneous with respect to 
        the driving timescale.
        Non-locality is thus implicitly enforced in computer
simulations,
        where the evolution of a single site 
        depends on the state of the entire system.
\bibitem{nota1}
        The relation $<s>\sim L^2$ has been already obtained from
        numerical simulations \cite{grasma,manna1} and proved 
        analytically in d=2 \cite{dhar}.
\end{thebibliography}
\end{document}